\documentclass[journal=aaemcq,manuscript=article,margin=0.5in]{achemso}

\usepackage[utf8]{inputenc}
\usepackage[version=3]{mhchem}      
\usepackage{amsmath}                
\usepackage{graphicx}               
\usepackage{chemfig}                
\usepackage{float}                  
\usepackage{placeins}               
\usepackage{tikz}                   
\usepackage[font=small,labelfont=bf]{caption}  
\usepackage{indentfirst}            
\usepackage{textcomp}               
\usepackage{ragged2e}               
\justifying
\usepackage{gensymb}                
\usepackage{ifthen}                 
\usepackage{calc}                   
\usepackage{refcount}              

\newlength{\maxfigwidth}
\newcommand{\fig}[4][\textwidth]{%
  \begin{figure}[H]
    \centering
    \includegraphics[width=\ifdim#1>\maxfigwidth \maxfigwidth \else #1\fi]{#2}
    \caption{#3}
    \label{fig:#4}
  \end{figure}
}

\newcounter{supfig}
\renewcommand{\thesupfig}{S\arabic{supfig}}

\newcommand{\supfig}[4][\textwidth]{%
  \stepcounter{supfig}
  \begin{figure}[H]
    \centering
    \includegraphics[width=\ifdim#1>\maxfigwidth \maxfigwidth \else #1\fi]{#2}
    \renewcommand{\thefigure}{\thesupfig}
    \caption{#3}
    \label{fig:sup#4}
  \end{figure}
}

\newcommand{\figref}[2][]{%
  {Figure~\ref{fig:#2}#1}%
}

\newcommand{\ubar}[1]{\overline{#1}}
\newcommand{\dc}{$d_C$ }

\author{Dafna Amichay}
\author{Alon Herman}
\author{Keren Shushan Alshochat}
\author{Eden Grossman}
\author{Baruch Hirsch}
\affiliation[]
{School of Electrical and Computer Engineering, Tel Aviv University, Tel Aviv, Israel.}
\author{Anchal Vashishtha}
\affiliation
{Department of Chemical Engineering, Ben-Gurion University of the Negev, Be'er-Sheva 8410501, Israel.}
\alsoaffiliation
{Ilse Katz Institute for Nanoscale Science and Technology, Be'er-Sheva 8410501, Israel.}
\author{Eran Edri}
\affiliation
{Department of Chemical Engineering, Ben-Gurion University of the Negev, Be'er-Sheva 8410501, Israel.}
\alsoaffiliation
{Ilse Katz Institute for Nanoscale Science and Technology, Be'er-Sheva 8410501, Israel.}
\author{Brian A. Rosen}
\affiliation{Department of Materials Science and Engineering, Tel Aviv University, Tel Aviv, Israel.}
\author{Gideon Segev} 
\affiliation[]
{School of Electrical and Computer Engineering, Tel Aviv University, Tel Aviv, Israel.}
\email{gideons1@tauex.tau.ac.il}

\title{Tuning electrochemical reactions with ratchet-based ion pumps}


\begin{document}

\begin{abstract}
 Electrochemical reactions are highly sensitive to the physical and chemical environment near the electrodes. Thus, controlling the electrolyte ionic composition and the electrochemical potential of specific ions can modify the overpotential of electrochemical reactions and enhance their selectivity toward the desired products. Ratchet-based ion pumps (RBIPs) are membrane-like devices that utilize temporal potential modulation to drive a net ionic flux with no associated electrochemical reactions.  RBIPs were fabricated by coating the surfaces of nanoporous alumina wafers with metals,  forming nanoporous capacitors. Placing the RBIP between two electrolyte compartments and applying an alternating signal between the metal layers resulted in a voltage buildup across the membrane, leading to ion pumping. Here, we demonstrate that by modifying the electrochemical potential of ions, RBIPs can accelerate or inhibit electrochemical reactions on the surface of adjacent water-splitting electrodes according to the RBIP input signal. Proton pumping towards a water-splitting cathode prevented proton depletion due to the hydrogen evolution reaction and maintained the pH in the cathode compartment. The combination of ion pumping and ion selectivity can enable the electrolyte composition to be tuned near the electrodes, providing greater control over the electrochemical process.
\end{abstract}

\section{Introduction}
Electrochemical reactions are sensitive to the ionic environment in the vicinity of the electrode. Thus, controlling the local ionic environment near electrodes can tune the reaction overpotential and current.  
\cite{Bard2001ElectrochemicalApplications,Kumunda2021ElectrochemicalReview,Resasco2018EffectsCO2,RodriguesPinto2024ElectrolyteElectrode,Horwitz2024TheFramework,Cui2023ContributionConfiguration,Lamoureux2019PHFirst-Principles}
For example, increasing the proton concentration near a cathode driving the hydrogen evolution reaction (HER)  can lower the reaction overpotential and enable higher currents and efficiency.
\cite{Cui2023ContributionConfiguration,Zheng2016UniversalEnergy,Lamoureux2019PHFirst-Principles,Zhang2024ProbingEC-SERS}
Furthermore, local control of the proton concentration near the cathode allows the HER to be driven without compromising the reaction kinetics, while allowing a higher pH of the solution in other parts of the system. Alternatively,  decreasing the proton concentration near the cathode can inhibit the HER in reactions in which hydrogen poisoning is to be avoided.
\cite{Chen2021CompetingFixation,Zheng2023TailoringEvolution,Tan2022EngineeringReaction,Guo2023DirectCatalyst,Wang2019AnomalousIons}
In multi-product reactions, pumping specific ions away from the working electrode can remove unwanted intermediate species, hinder a competing reaction, and provide another handle for controlling the reaction selectivity.
For example, pumping protons away or towards a \ce{CO2} reduction cathode can help tune the \ce{H2} to \ce{CO}  ratio in \ce{CO2} reduction systems.
\cite{Saha2022SelectivityCO2Reduction,Ummireddi2021InhibitionElectrocatalysts}

Ratchet-based ion pumps (RBIPs) utilize the temporal modulation of the electric potential to drive a non-zero-time averaged ionic current with no associated electrochemical reactions.\cite{Kautz2025ASeparations} The RBIPs are fabricated by coating the two surfaces of nanoporous anodized aluminum oxide (AAO) wafers with thin metal layers, forming nanoporous capacitor-like structures. When placed as a membrane between two electrolyte compartments, the non-linear capacitance of the electrode double layers results in a dispersion of the charging and discharging time constants at each RBIP surface. This leads to a buildup of an electric potential difference across the RBIP membrane and to a net ion flux through the RBIP.
In the first experimental demonstration, the RBIP induced ionic currents in the order of $10{\mu A}{cm^{-2}}$ and a voltage of about $80 mV$. The RBIP also showed a noticeable output for signals with amplitudes as low as 50 mV (peak-to-peak), indicating that ion pumping is not carried out by redox reactions. Moreover, RBIP-driven electrodialysis was demonstrated, reaching a $50\%$ decrease in conductivity in a dilution cell.\cite{Kautz2025ASeparations} 
Theoretical studies have shown that RBIPs can drive selective ion separation by transporting ions with the same charge in opposite directions according to their diffusion coefficients, or drive ambipolar transport in which both cations and anions are transported in the same direction.
\cite{Herman2024AmbipolarMembranes,Herman2023Ratchet-BasedSeparations} 
In this work, we demonstrate how RBIPs can tune the overpotential and current of electrochemical reactions by pumping protons toward or away from water-splitting electrodes. By directing protons toward a Pt cathode, the RBIP enhanced the hydrogen evolution reaction (HER) and compensated for the proton depletion that resulted from the reaction. Alternatively, by pumping ions away from the water-splitting cathode, the RBIP enhanced proton depletion and increased the pH in the cathode compartment.
The introduction of selective ion pumping membranes into electrochemical systems can enable the control of the overpotential of electrochemical reactions and fine-tune more complex reactions, thereby providing an additional degree of freedom for the electrochemical process.

\section{Experimental}
\subsection{Sample fabrication}
Anodized Aluminum Oxide (AAO) wafers (60 nm pore diameter and 50 microns thickness, InRedox LLC) were annealed at $\ 650^{\circ} C  $ for 10 hours.\cite{Choudhari2012FabricationMembranes} Then, a 40-50 nm thick (planar equivalent) gold thin film  was deposited on each surface using magnetron sputtering.
Last, both surfaces were coated with an 8 nm layer of \ce{TiO2} or \ce{Al2O3,} using Atomic Layer Deposition (ALD). 
The \ce{TIO2} ALD process was as described by Vega et al.\cite{Vega2017DiffusiveOxides}  The exposure time to the precursors was set to 1 s and the purging time was set to 5 s.
Atomic layer deposition of \ce{Al2O3} was carried out using a Gemstar XTTM tabletop ALD
system. The chamber pressure was approximately 170 mTorr, and Trimethyl aluminium
(TMA) and \ce{H2O} were used as precursors. The process comprised of alternating pulses of
TMA and  \ce{H2O}, with the pulse length of 250 ms and 150 ms, respectively. The expo value was
opened for 60 s after each pulse and after that Ar gas at a flow rate of 10 SCCM was purged
to remove unreacted precursors from the chamber. To achieve a thickness of ~8-10 nm, 60 deposition cycles were done.
\cite{Correa2015ChemicalDeposition,George2010AtomicOverview}

\subsection{Experiment design}

All experiments were conducted in a two-compartment PEEK electrochemical cell as shown in \figref[a]{systemsetup}.
The RBIP was placed as an active membrane separating the two compartments.
\figref[b]{systemsetup} shows an illustration of the experimental setup. 
A Pt working electrode with a 1.6 mm diameter working area (ALS Japan 002313, denoted WE) was placed in one electrolyte compartment. A Pt wire counter electrode (ALS Japan 002233, denoted CE)  and a reference electrode (Ag/AgCl in saturated NaCl, ALS Japan 013393, denoted RE) were placed in the other compartment. The electrode potential or current was controlled using a potentiostat (Zhaner ZENNIUM X).
The RBIP was connected to a signal generator, which provided the ratchet input signal (Keysight 33500B).
In pH regulation experiments, pH measurements were taken using a Mettler Toledo micro nano pH electrode. Every 30 minutes, three separate pH measurements were taken and averaged.
The solution was refreshed after each experiment.
When changing samples, the cell was cleaned using the following process: the PEEK parts were sonicated in IPA for 15 minutes, followed by a sonication in distilled water.
Next, the cell was immersed in \ce{HNO3} (20\%) for 2 hours.
Finally, the cell was sonicated in distilled water for 15 minutes, 3 times.
The working electrode was cleaned using a commercial polishing kit (ALS Japan), Alumina polishing paper, and an alumina slurry solution. It was then sonicated in distilled water for 5 minutes.

\subsection{RBIP performance characterization}
The RBIP performance was characterized in response to periodic square wave input signals:
\begin{equation}
V_{in}(t)=\left\{
\begin{array}{rl}
V_{p-p}/2, & 0<t<=d_cT\\
-V_{p-p}/2, & d_cT<t<=T
\end{array}\right.
\label{eq:Vin}
\end{equation}
where $V_{p-p}$ is the input signal amplitude (peak to peak), $T$ is the input signal temporal period, and  \dc is the input signal duty cycle- the ratio between the time the input signal is at its high value and its temporal period. To characterize the RBIP performance, the working electrode current or potential was measured in response to the RBIP input signal.
These measurements are referred to as duty cycle sweeps, and they can be conducted in chronoamperometry mode to measure the temporally averaged RBIP-induced current, $\ubar{I}_{out}$, or chronopotentiometry mode to measure the RBIP-induced voltage, $\ubar{V}_{out}$.
In chronoamperometry duty cycle sweeps, the working electrode (WE) potential is fixed, and its current is measured while varying the input signal duty cycle.
In chronopotentiometry duty cycle sweeps, the working electrode current is fixed, and the WE potential is measured while the input signal duty cycle is varied. 
For each duty cycle, a square wave input signal was applied to the RBIP for $t_{on} =$ 30 s and then ${V}_{in}$ was set to 0 V for $t_{off} = 30$ s.
The temporally averaged RBIP-induced current (voltage) follows.
	\begin{equation}
	    \ubar{I}_{out}= \ubar{I}_{ON}-\ubar{I}_{OFF} 
        \label{eq:iout}
	\end{equation}
Where $\ubar{I}_{ON}$ is the temporally averaged WE current (potential) measured when the RBIP is ON, and $\ubar{I}_{OFF}$   is the temporally averaged WE current (potential) when the RBIP is OFF:
\begin{equation}
	\ubar{I}_{ON}=\frac{1}{t_{AV}}  \int_{\frac{t_{ON}-t_{AV}}{2}}^{\frac{t_{ON}+t_{AV}}{2}} 
    I_{ON}(t)\, dt     
      \label{eq:ion}
\end{equation}
\begin{equation}
	\ubar{I}_{OFF}=\frac{1}{t_{AV}}  \int_{\frac{t_{OFF}-t_{AV}}{2}}^{\frac{t_{OFF}+t_{AV}}{2}} I_{OFF}(t)\, dt 
      \label{eq:ioff}
\end{equation}
$I_{ON}$  is the current measured when the RBIP was ON, $I_{OFF}$ is the current measured when the RBIP was OFF,  $t_{AV}=$ 20 s is the length of the temporal window in which the output is averaged.

\section{Results and discussion}
\subsection{Electrochemical characterization.}
RBIP samples were fabricated by coating both sides of AAO nanoporous wafers (pore diameter of 60 nm) with 50 nm thick (planar equivalent) gold thin films, followed by an 8 nm thick \ce{TiO2} or \ce{Al2O3} layers deposited with atomic layer deposition. For more details on the fabrication process, refer to the experimental section. The RBIP was placed as an active membrane between two compartments of an electrochemical cell. A platinum working electrode was placed in one compartment, and a counter electrode and a reference electrode were placed in the opposite compartment. The working, reference, and counter electrodes were connected to a potentiostat, and the two RBIP metal surfaces were connected to a signal generator. \figref[a]{systemsetup} shows a photograph of the electrochemical cell used for the measurements, and \figref[b]{systemsetup} shows a schematic illustration of the experimental setup. The $R^+$ compartment in the electrochemical cell is the compartment adjacent to the RBIP surface connected to the positive lead of the signal generator, and the $R^-$ compartment is the compartment adjacent to the RBIP surface connected to the negative lead of the signal generator.

\fig[15cm]{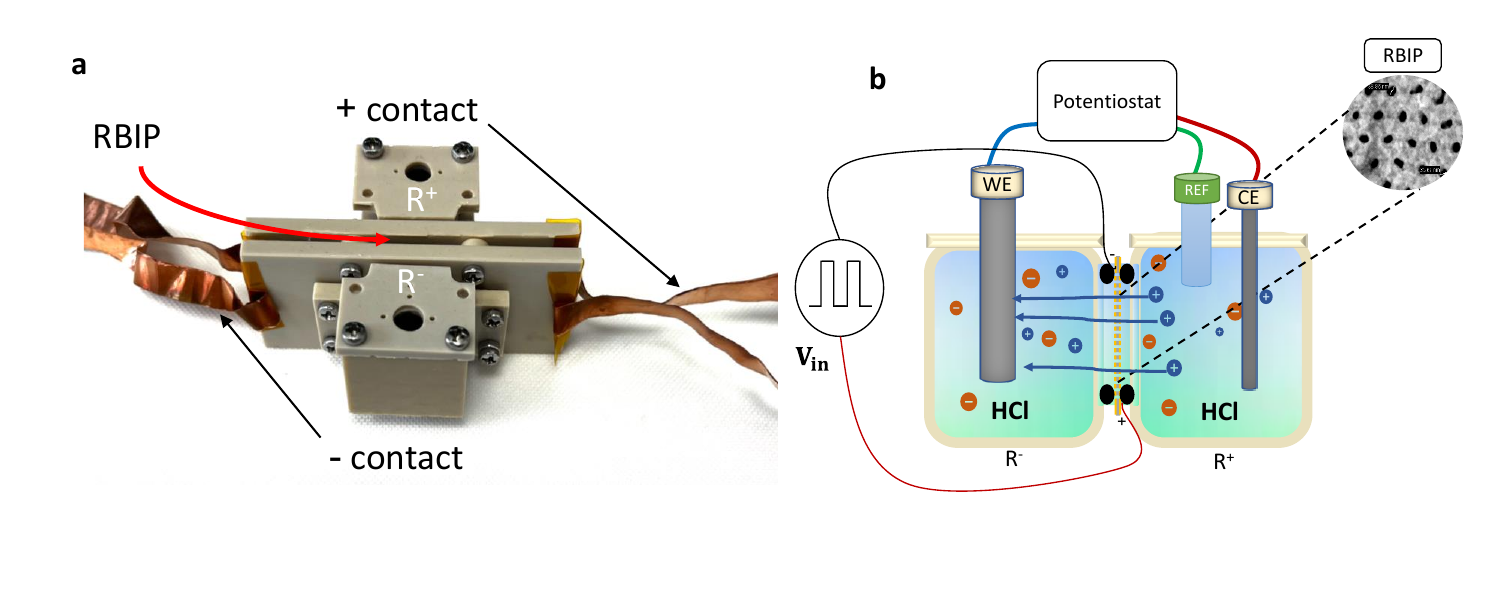}{(a) A photograph of the electrochemical cell used for RBIP characterization. The RBIP is placed between the two compartments, and the copper tape is used to contact the RBIP. (b) A schematic illustration of the experimental setup.  The RBIP is placed as a membrane separating the two compartments of an electrochemical cell.  A signal generator connected to the RBIP provides the ratchet input signal $V_{in}$.
The working, reference, and counter electrodes (WE, REF, and CE, respectively) are connected to a potentiostat.}{systemsetup}
 
To ensure that the system is chemically clean and stable, cyclic voltammetry (CV) measurements of the platinum working electrode were conducted. The CV measurements were performed in \ce{HCl} (pH=4.2, 0.2m\textsc{m} and pH=2.56, 2.75m\textsc{m}), \ce{KCl} (pH=6.2, 0.2m\textsc{m}),  and \ce{H2SO4} (pH=2.5, 1.6m\textsc{m}) aqueous solutions. All measurements were taken in a PEEK electrochemical cell with a 3-electrodes setup  (\figref[a-b]{systemsetup}) at a scan rate of 50 mVs$^{-1}$. More details on the setup can be found in the experimental section.
\figref{allCVpre}  shows the measured voltammograms. Chemical processes were assigned to each peak by comparing the CV curves to those of well-studied Pt electrodes in \ce{H2SO4} and to HCl aqueous solutions.\cite{Daubinger2014ElectrochemicalStudy,Jerkiewicz2004Surface-oxideMeasurements}   The cyclic voltammetry peaks of \ce{HCl} at pH=2.56 closely align with the \ce{H2SO4} peaks, and the electrochemical water windows of both voltammograms are similar. Hydrogen under-potential deposition (HUPD) occurs when protons are absorbed to the cathode at potentials that are more positive than the equilibrium reaction voltage.
  \cite{Daubinger2014ElectrochemicalStudy,Zolfaghari1997EnergeticsSpecies,Dubouis2019TheDescriptors,Yang2019UnderstandingPlatinum,Deng2022AcceleratedActivity,Sheng2015CorrelatingEnergy}  The arrows in \figref{allCVpre} point to the HUPD peaks.
 The CV curves measured with  \ce{HCl}  pH=4.2 and \ce{KCl} show  HUPD peaks that are less defined than the peaks in \ce{H2SO4} or \ce{HCl} at a pH of 2.56. The HUPD peaks in pH= 4.2   \ce{HCl}  and \ce{KCl}  are indicated by dashed arrows. 

\fig[12cm]{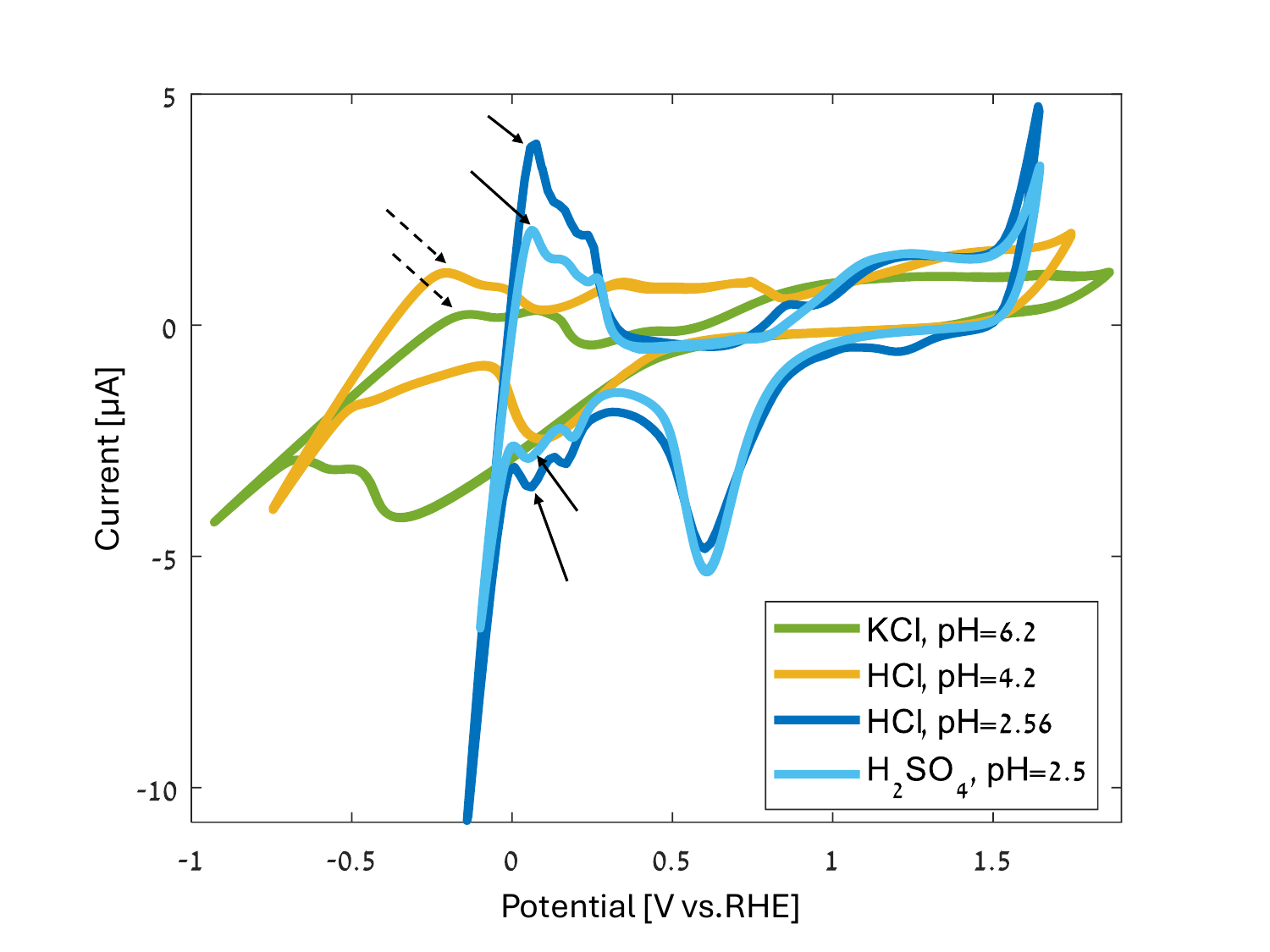}{Cyclic voltammetry measurements of the platinum working electrode. The scan rate  is 50 mVs$^{-1}$, and the solutions are aqueous  \ce{HCl} (pH=4.2 ,0.2 m\textsc{m} and pH=2.56, 2.75 m\textsc{m} ),  \ce{KCl} (pH=6.2, 0.2 m\textsc{m}), and \ce{H2SO4} (pH=2.5, 1.6 m\textsc{m}). The arrows point to the HUPD peaks in each solution.}{allCVpre}

\subsection{Current enhancement }

The performance of the RBIP was assessed by conducting a chronoamperometry duty cycle sweep in which the working electrode (WE) potential was set to 0 V vs. RHE, and the WE current was measured.
 The sample was fabricated as described in the experimental section with a \ce{TiO2} ALD coating.
The electrolyte was a 0.2 m\textsc{m} HCl aqueous solution, the working electrode was placed in the $R^-$ compartment, and the counter and reference electrodes were placed in the $R^+$ compartment (this electrode configuration is denoted hereafter as configuration A).
The input signal was set to 0 V for 30 seconds (RBIP OFF), after which a square wave input signal was applied for 30 seconds (RBIP ON).
The input signal frequency was 100 Hz, and the amplitude was $V_{p-p}= 1.4\space V$. \figref[a]{constvoltdc} shows the current measured during the duty cycle sweep.  The shaded regions mark the times when the RBIP was off, and the white regions indicate the times when a square wave input signal was applied.
The color bar indicates the duty cycle of the input signal when the RBIP was ON.
\figref[c]{constvoltdc} shows the current measured during a chronoamperometry duty cycle sweep in which the WE was in the $R^+$ compartment and the reference and counter electrodes are in the $R^-$ compartment (see illustration in the inset of \figref[c]{constvoltdc}, this configuration is denoted as configuration B). The RBIP-induced current is the time-averaged current measured when the input signal is applied to the RBIP, reduced by the time-averaged current measured while the RBIP was OFF (more details on the calculation of the RBIP-induced current can be found in the experimental section). 
\figref[b]{constvoltdc}  and \figref[d]{constvoltdc}  show the RBIP-induced current as a function of the duty cycle when the system was in configurations A and B, respectively.
During the HER, protons are directed toward the cathode. However, the RBIP action alters this proton flux, accelerating or hindering it, depending on the cathode's position (configuration A or B) and the characteristics of the input signal.
When a constant bias is applied (i.e., a duty cycle of 0 or 1), the RBIP-induced current is negligible and diminishes rapidly.
At low duty cycles (\dc $ <0.5$), the RBIP drives a more cathodic current with an increase of up to $1\mu A$ in configuration A, and a less cathodic current in configuration B. 
Conversely, at high duty cycles (\dc$ >0.5$), the RBIP  drives a more cathodic current in configuration B, and a less cathodic current in configuration A.
Thus, for \dc $ <0.5$ the RBIP drives protons towards the $R^-$ compartment, and for \dc$ >0.5$ the RBIP drives protons towards the $R^+$ compartment. Hence, at each duty cycle, the RBIP exerts a force on the protons in a direction independent of the flux induced by the electrochemical reactions at the working and counter electrodes.
The anodic currents in configuration A at duty cycles above 0.5  (\figref[a]{constvoltdc}) are a result of the RBIP inducing a voltage of several dozens of mVs, which shifted the working electrode operating point to the potential of the HUPD anodic peak. In prior experimental demonstrations of RBIPs, the sign of the output did not change with the duty cycle. \cite{Kautz2025ASeparations} However, in \figref[b]{constvoltdc}  and \figref[d]{constvoltdc}, the RBIP output is approximately anti-symmetric with respect to a duty cycle of 0.5, indicating that in this sample, the two surfaces showed a similar non-linear capacitance. \cite{Kautz2025ASeparations} This may be a result of a change in charge distribution within the pore and in the resting potential of the electrodes in this specific solution and concentration.  Engineering spatially asymmetric devices will further increase their output and determine the ion pumping direction.
\cite{Kautz2025ASeparations} Nevertheless, the ratchet effect on the electrochemical reaction is always consistent: pumping protons toward the cathode reduces the overpotential and increases the current. 

\fig[15cm]{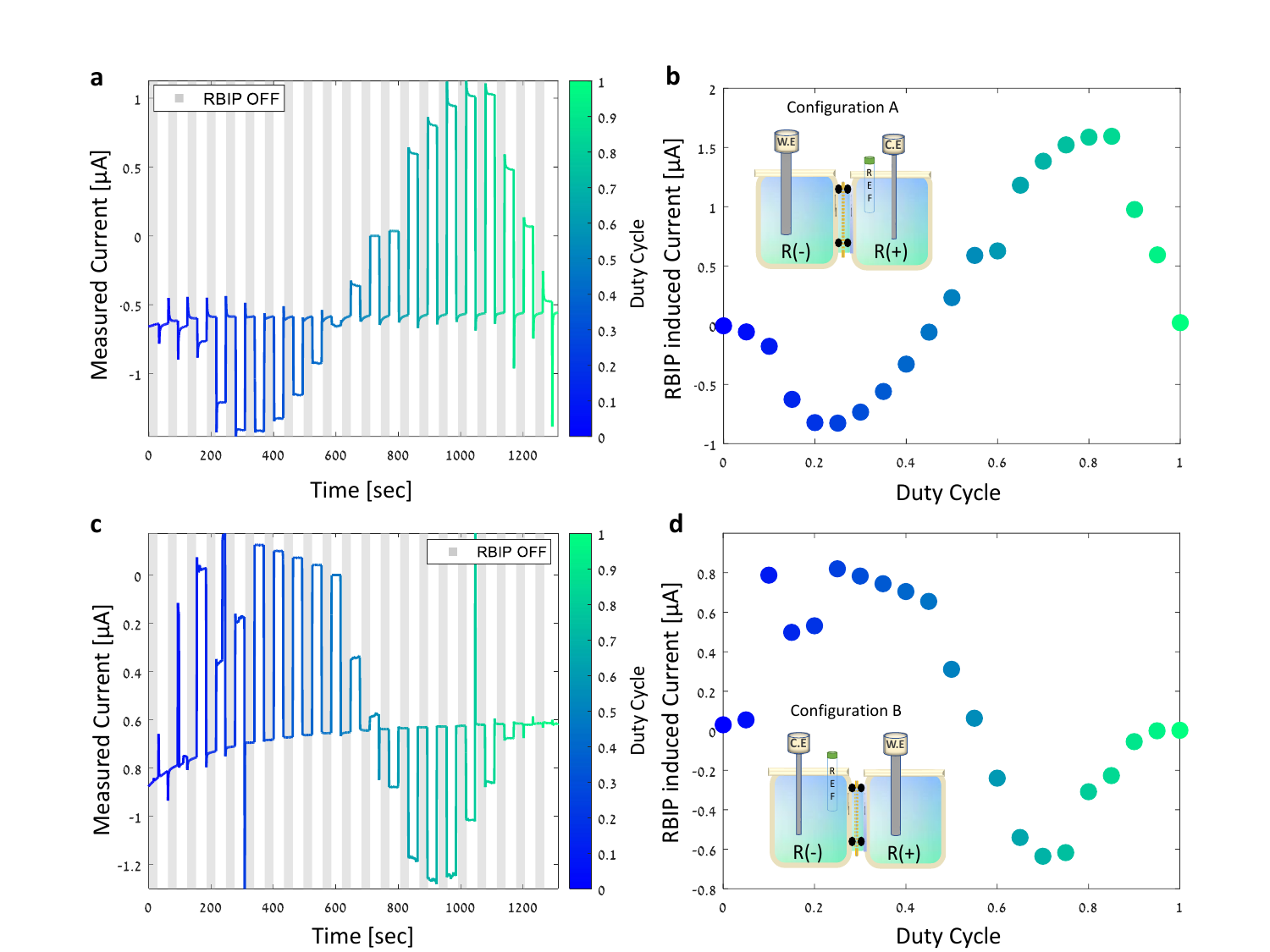}{(a, c) The measured current for a system in configuration A and B, respectively. The shaded areas in a and c indicate the times when the RBIP is OFF ($V_{in}$ = 0 V), and the bright areas indicate the times when it is on.
The colorbar indicates the duty cycle of the input signal when the RBIP is ON.
The duration of each ON and OFF cycle was 30 seconds.
(b, d) The temporally averaged RBIP-induced current as a function of the duty cycle obtained from (a, c).
In all measurements, the WE potential  is 0 V vs. RHE. The input signal frequency is 100 Hz, and the amplitude is $V_{p-p}= 1.4\space V$. The sample was fabricated as described in the experimental section with a \ce{TiO2} ALD coating. 
The compartments are filled with a 0.2 m\textsc{m} HCl aqueous solution.
The insets in b and d are illustrations of configurations A and B, respectively.}{constvoltdc}
\pagebreak
\subsection{pH regulation }
Next, we show how RBIPs can help regulate the pH in the cathode compartment. To do so, the system was operated in configuration A at a constant working electrode current of $-3\space \mu A$ . The pH of both electrolyte compartments was measured during operation. 
First, a baseline measurement was performed with the RBIP disconnected. 
The RBIP was then operated with a duty cycle of $0.2$, a frequency of 100 Hz, and an amplitude of $V_{p-p}= 1.4\space V$.
Lastly, the RBIP was operated with a duty cycle of $0.6$ and the same frequency and amplitude.
The sample was fabricated as described in the experimental section with a \ce{TiO2} ALD coating.
\figref{phmeas} shows the pH of the two compartments measured in the three experiments.
The RBIP pore walls are positively charged, leading to a partial permselectivity that impedes proton transport through the RBIP. \cite{Petukhov2017LiquidMembranes,Parks1962TheOxides}
When HER was driven at a current of $-3\space \mu A$, proton consumption by the cathode was faster than proton transport through the RBIP.  As a result, the proton concentration in the cathode compartment was reduced and the pH increased.
Driven with a duty cycle of $0.2$ , the RBIP pumped protons toward the cathode. The augmented proton flux compensated for the proton consumption by the electrochemical reaction and maintained a more moderate pH in the cathode compartment.
However, at a duty cycle of $0.6$ the RBIP pumped protons away from the cathode compartment, increasing proton depletion and resulting in a pH higher than the baseline.
The change in pH in the anode compartment is within the error of the measurement system.
The change in pH in response to the ratchet action demonstrates that RBIPs can regulate the electrolyte composition and pH in electrochemical systems. Repeating this experiment with a different sample and testing both electrode configurations A and B again demonstrates the directionality of the RBIP, as shown in Supplamentary Figure S1. The direction of ion pumping determines whether the RBIP enhances or mitigates proton depletion in the cathode compartment (Supplementary Information, Section 1).
In electrochemical systems with multiple competing reactions, the RBIP can optimize the reaction selectivity. For instance, extracting protons from the cathode compartment  can minimize hydrogen generation in \ce{CO2} reduction systems, where HER is a competitor.\cite{Marcandalli2022ElectrolyteCO,Saha2022SelectivityCO2Reduction} 
Ion pumping, as demonstrated above, can pave the way towards local pH control. For example, if operating with a small distance between the RBIP and the cathode, the RBIP can increase the proton concentration locally near the cathode, thus facilitating HER without compromising reaction kinetics while maintaining a more moderate bulk electrolyte pH. 
\fig[\textwidth]{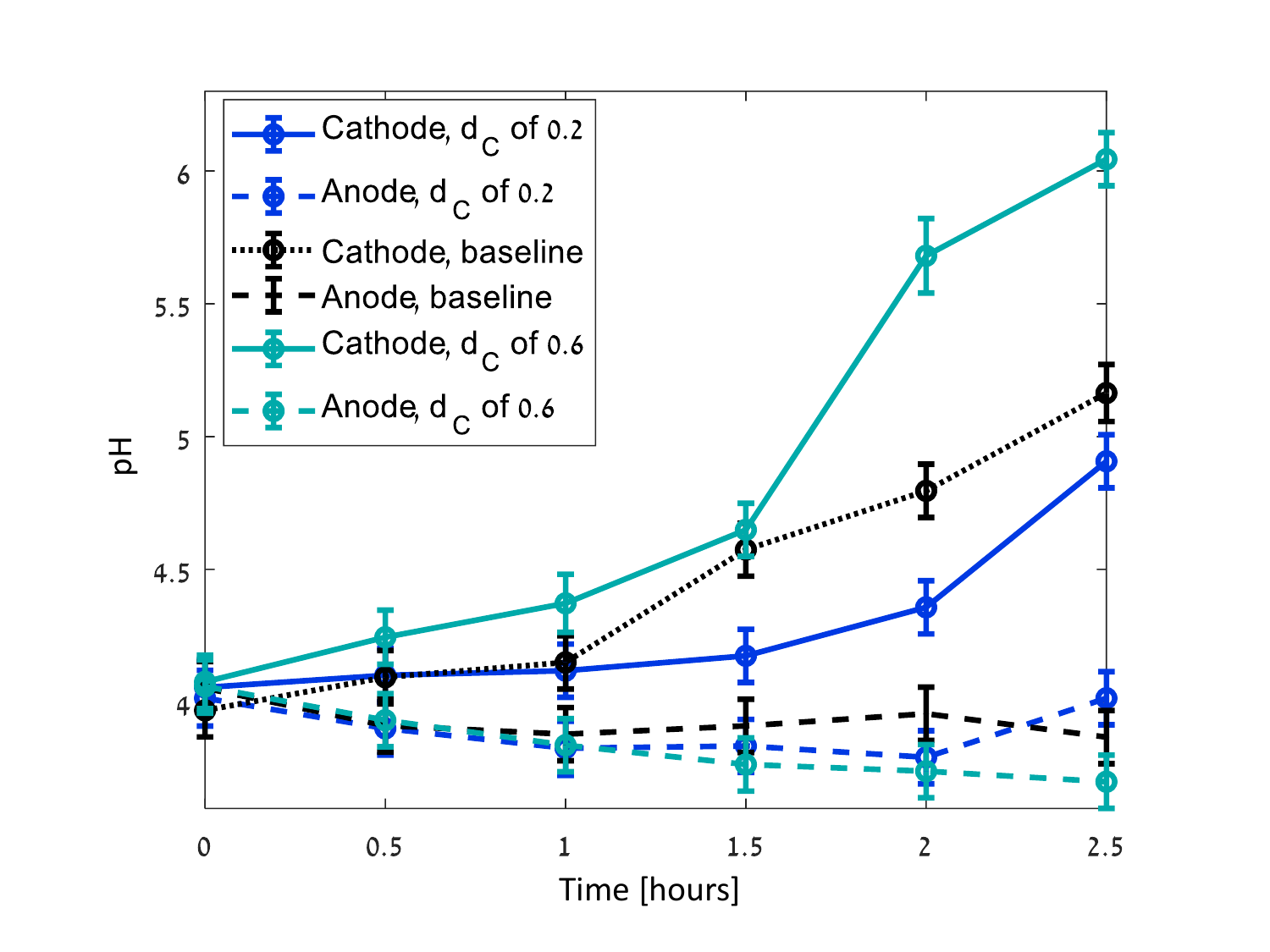}{The pH of the anode and cathode compartments during 2.5 hours of operation with the RBIP operating and when it is disconnected. The cathode current is $-3\space\mu  A$, and the system is in configuration A.
The sample was fabricated as described in the experimental section with a \ce{TiO2} ALD coating. 
The compartments are filled with a 0.2m\textsc{m}  \ce{HCl} aqueous solution, the input signal frequency is $100 Hz$, and the amplitude is $V_{p-p}= 1.4\space V$.}{phmeas}

\pagebreak
\subsection{Voltammogram shift}
The RBIP's effect on the electrochemical system was measured by comparing CV measurements with the RBIP driven with various duty cycles. The cell was filled with a 2.6 m\textsc{m} HCl aqueous solution. The sample was fabricated as described in the experimental section with an alumina  ALD coating. The scan rate was 50 mVs$^{-1}$. First, a voltammogram was taken with the RBIP OFF ($V_{in} = 0 V$). Then, a CV measurement was carried out with a square wave input signal applied to the RBIP.  Last, the input was set again to 0 V, and a cyclic voltammetry was remeasured. This procedure was repeated with the  duty cycle varied between 0 and 1 in steps of 0.1. The frequency was 15 kHz, and the amplitude was $V_{p-p}= 1.4\space V$. 
\figref[a]{CVsweepmin} shows the measured voltammograms. The black dashed curves are voltammograms measured while the RBIP was OFF before and after each time the RBIP was ON, and the colored curves are voltammograms measured while various input signals were applied to the RBIP. The color coding corresponds to the input signal duty cycle. 
The RBIP operates as a voltage source, adding (or subtracting) to the potential applied by the potentiostat. As a result, the peaks and current onsets of the CV curves shift by the RBIP-induced voltage. For duty cycles below 0.5, the RBIP induced a cathodic voltage that drove protons toward the working electrode. As a result, the HER current onset shifted anodically. However, for duty cycles above 0.5, the RBIP drove protons away from the working electrode, and the HER current onset shifted to more cathodic potentials. Conversely, the OER onset shifted cathodically for duty cycles above 0.5 and shifted anodically for duty cycles below 0.5. Constant biasing of the RBIP (duty cycle of $0$ and $1$) did not affect the CV curve, as no voltage develops across the RBIP. 
The voltammograms with the ratchet OFF ($V_{in}=0 \space V$) overlay almost perfectly, indicating minimal changes to the RBIP, solution, and electrodes during operation.
 We define the HER onset potential as the potential of the local cathodic current minimum (in terms of absolute value) just before the sharp increase in the cathodic HER current (noted (i) in \figref[a]{CVsweepmin}). The proton desorption peak potential and current are the potential and current at which the curve reaches an anodic current maximum after driving HER ((iii) in  \figref[a]{CVsweepmin}). \figref[b]{CVsweepmin} shows the potentials of the  HER onset and the proton desorption peak extracted from the CV curves in \figref[a]{CVsweepmin} as a function of the input signal  duty cycle. At duty cycles below 0.5, the voltage induced by the RBIP contributed to the HER. The proton desorption peak and HER onset potential reached potentials as high as 0.236 and 0.087 V vs. RHE, respectively. At  input signal duty cycles above 0.5, the RBIP induced a voltage that negates driving HER at the working electrode. As a result, the WE potential must be more cathodic to drive the same reaction, and the proton desorption peak and HER onset potentials reached -0.07 and -0.137 V vs. RHE, respectively. The effect of the RBIP on the two potentials is very similar, indicating that in this configuration, the RBIP acts as a voltage source that is added to the potential applied by the potentiostat. This was verified by comparing the CV curves measured with the ratchet ON to CV curves measured in a potential range that is shifted by a magnitude similar to the voltage induced by the RBIP (see supplementary information section 2 for more details).
\figref[c] {CVsweepmin} shows the ratchet-induced current extracted from points (ii) and (iii) on the CV as a function of the input signal duty cycle (Figure S6 shows how the proton desorption peak ratchet-induced current is extracted). The voltage induced by the ratchet led to a shift in the HER current onset and to a corresponding change in the HER and proton desorption currents. The resulting HER current change reached $9\mu A$  and $8.7\mu A$  in absolute value for duty cycles of 0.4 and 0.6. The proton desorption current change reached  $4.9\mu A$ (in absolute value) for duty cycle of 0.4 and $5.4\mu A$ for duty cycle of 0.6.

\fig[\textwidth]{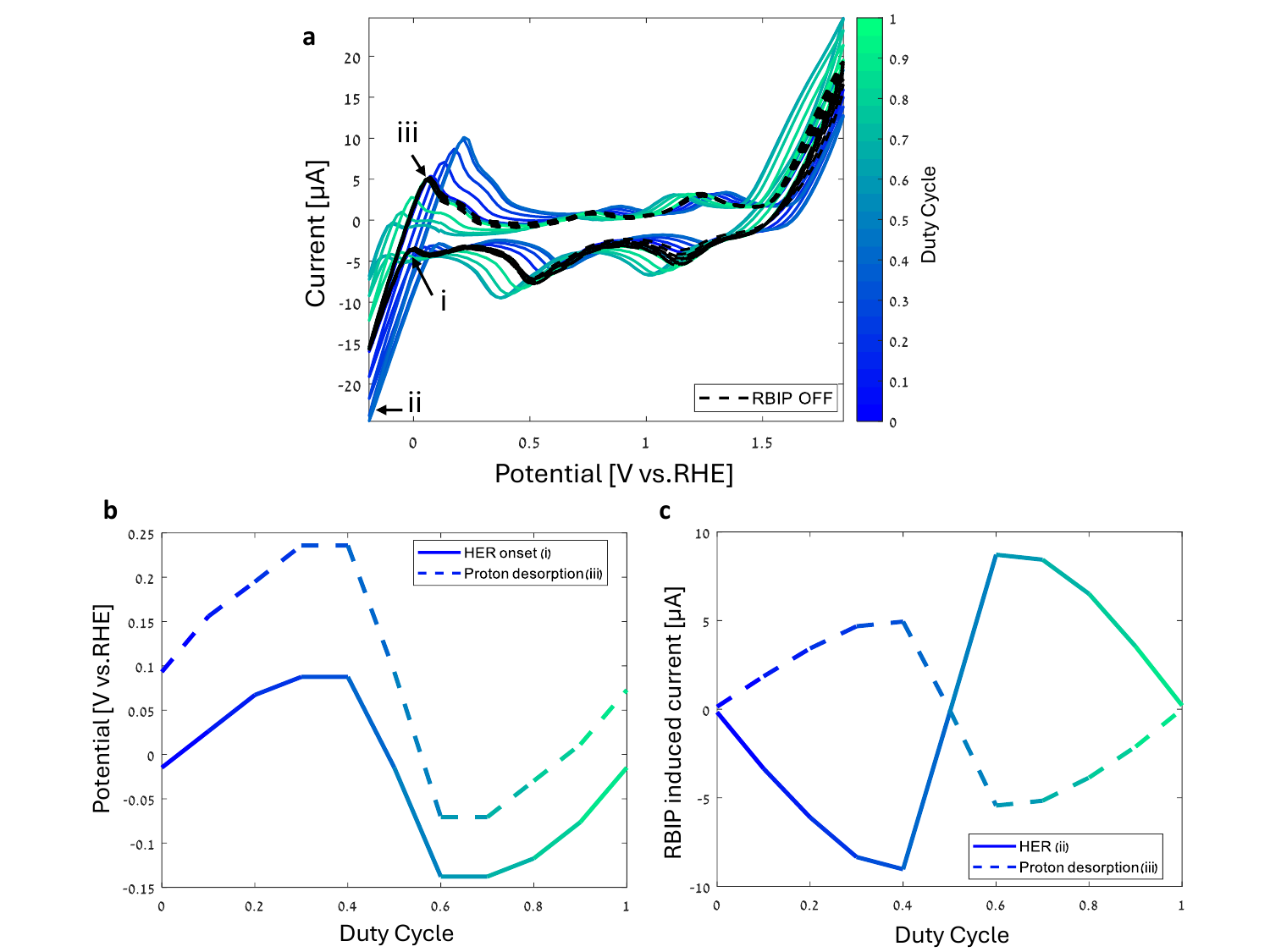}{(a) Three-electrode cyclic voltammetry measurements with the RBIP OFF ($V_{in}=0$, dashed black curves) and with the RBIP driven with several  input signal duty cycles. The scan rate is 50 mVs$^{-1}$. The sample was fabricated as described in the experimental section with an alumina ALD coating. The electrolyte is 2.75 m\textsc{m} \ce{HCl}  aqueous solution. The input signal frequency is 15 kHz, and the amplitude is $V_{p-p}= 1.4\space V$. (b)  The potential of the HER onset and proton desorption peak extracted from (a) as a function of the input signal  duty cycle. (c) The extracted RBIP-induced HER and proton desorption currents as a function of the input signal duty cycle. }{CVsweepmin}

\pagebreak
The RBIPs effect on the CV was also measured and analyzed in configuration B (Figure S3). For duty cycles below 0.5 the CV curves shifted anodically in configuration A (where the working electrode is in the $R^-$ compartment) and shifted cathodically in configuration B (in which the working electrode is in the $R^+$ compartment). Conversely, for duty cycles above 0.5, the CV curves shifted cathodically in configuration A, but shifted anodically in configuration B. 
This demonstrates that this sample induces a voltage that directs protons to the $R^-$ compartment when driven with a duty cycle below 0.5 and to the $R^+$ compartment with a duty cycle above 0.5.
This experiment was repeated with in a 1.6 m\textsc{m} \ce{H2SO4} aqueous solution, and showed the same trends with slightly higher output (an anodic shift of 175.9 mV). Thus, the effects observed are a result of an inherent RBIP functionality and are not specific to the solution chemistry (see supplementary information section 4 for more information). To assure that the observed ratchet action is not due to feedbacks introduced by the potentiostat, experiments were also conducted in a two-electrode arrangement. As with \figref[a]{CVsweepmin}, the introduction of an alternating signal to the RBIP resulted in cathodic or anodic shifts in the current onsets and peaks according to the input signal duty cycle.  However, the application of a constant bias to the RBIP had no effect on the CV. More information on these measurements can be found in the supplementary information section 5.
The effect of the RBIP on the current-voltage relationship can be viewed as that of a transistor where the voltage applied to one electrode (the gate) controls the current-voltage relationship between two other electrodes (source and drain). Thus, controlling the current between the working and counter electrodes by applying various signals to the RBIP can be utilized to obtain a transistor-like functionality. Combining this functionality with ion-ion selectivity\cite{Herman2023Ratchet-BasedSeparations} can lead to the development of highly controlled electrochemical systems where only specific reactants are allowed to reach the electrodes. Such functionality may be useful in applications such as precise drug delivery systems, and amplified ion-specific chemical sensors.
\pagebreak
\section{Conclusions}
Ratchet-based ion pumps were utilized as active membranes in an electrolysis cell. The application of an alternating input signal to the RBIP results in a buildup of a ratchet-induced voltage between the two sides of the membrane. This voltage is then utilized to accelerate or suppress electrochemical reactions on the surface of the working electrode according to the input signal duty cycle.  pH regulation was demonstrated by pumping protons  towards the water splitting cathode, thus compensating for proton depletion by the reaction. Conversely, at a higher input signal duty cycle, proton depletion was enhanced by pumping protons away from the cathode during the water splitting process.  Cyclic voltammetry measurements during the RBIP operation showed that the RBIP can shift the onset potential of electrochemical reactions by up to 142 mV, demonstrating an electrochemical transistor-like behavior.  The RBIP's ability to tune the onset potential of redox reactions and to regulate the chemical environment near electrodes can provide an added degree of freedom in electrochemical systems used for renewable fuels, chemical sensors, and other applications.

\begin{acknowledgement}
This work was partially funded by the Israeli Ministry of Energy. This work is partially funded
by the European Union (ERC, ESIP-RM, 101039804). Views and opinions expressed are
however those of the author(s) only and do not necessarily reflect those of the European
Union or the European Research Council Executive Agency. Neither the European Union
nor the granting authority can be held responsible for them. We acknowledge the
contribution of TAU Nano center for providing the sputtering and e-beam evaporation
equipment, and the HRSEM

\end{acknowledgement}

\begin{suppinfo}
Supporting Information is available from the author.

\end{suppinfo}

\bibliography{references}

\end{document}


\section{pH regulation in configurations A and B}
The experiment described in Figure 4 was repeated in configurations A and B with a different sample.
\supfigref[a]{phbothsides} shows the pH measured in the two electrode compartments while the RBIP was disconnected and when it was operating. 
\supfigref[b,c]{phbothsides} shows the ratchet-induced voltage measured in a Chronopotentiometry duty cycle sweep for configurations A and B, respectively.
In all these measurements, the working electrode current is $-3\mu A$, the compartments were filled with a 0.2m\textsc{m}  \ce{HCl} aqueous solution, the input signal frequency is $100 Hz$, and the amplitude is $V_{p-p}= 1.4\space V$. The sample was fabricated as the one as in Figure 4.
As shown in \supfigref[b,c]{phbothsides}, the RBIP-induced voltage shows a typical ratchet behavior  with low outputs at extreme duty cycles, and peaking at \dc=0.5 for both configurations. In configuration A, the RBIP-induced voltage is negative for all duty cycles, indicating that a higher overpotential is required in order to maintain the same HER current. Thus, the RBIP is pumping ions away from the cathode compartment at every duty cycle. In configuration B, the RBIP output is positive. Thus, a lower overpotential is required in order to maintain the reaction. Hence, in this case, the RBIP pumps ions towards the working electrode for all duty cycles. As observed in \supfigref[a]{phbothsides}, at a duty cycle of 0.7, when the system was in configuration A,  protons were pumped away from the cathode compartment, and the pH rose above the baseline. In contrast, when the system was in configuration B (switching the position of the working electrodes), protons were pumped toward the cathode compartment, maintaining the pH closer to its initial value. 
\supfig[\textwidth]{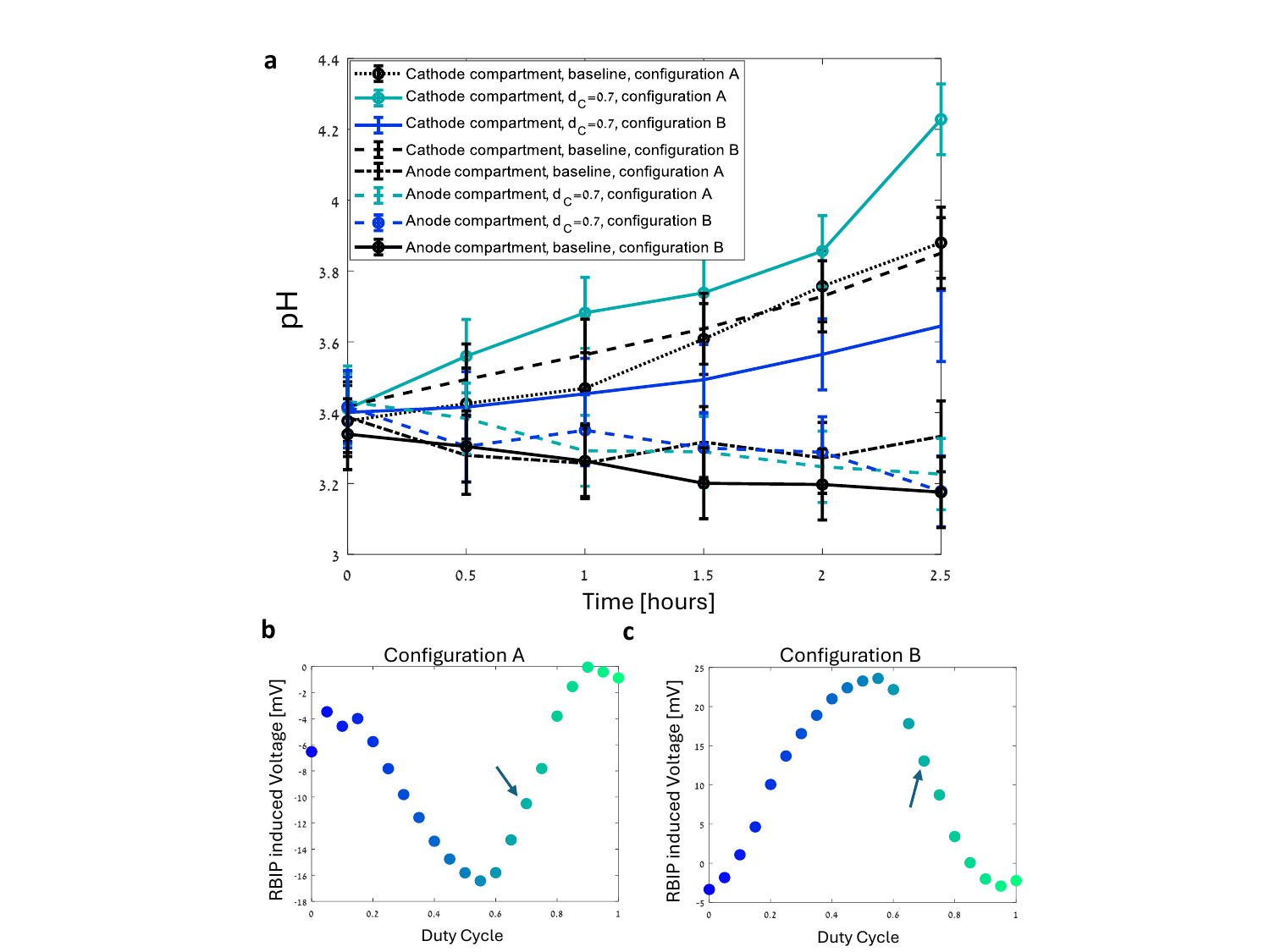}{(a) The measured pH in the anode and cathode compartments when driving water splitting in configurations A and B, with the RBIP operating and when the RBIP was off ($V_{in} =$ 0 V).  (b-c) The ratchet-induced voltage measured in a Chronopotentiometry duty cycle sweep for configurations A and B, respectively. The working electrode current is $-3\mu A$ the compartments were filled with a 0.2m\textsc{m} \ce{HCl} aqueous solution, the input signal frequency was $100 Hz$, and the amplitude is $V_{p-p}= 1.4\space V$. 
The sample was fabricated as described in the experimental section with a  \ce{TiO2} ALD coating.}{phbothsides}
\pagebreak
\section{Cyclic voltammetry shift}
The RBIP functions as a voltage source connected in series to the potentiostat, thus adding or subtracting from the voltage applied to the working electrode during the CV measurement. This was verified by comparing CV curves measured with the RBIP ON with CV curves measured with RBIP OFF, yet at a potential range that is shifted by the RBIP-induced voltage.
First, CV  curves were measured in the potential range between -0.342 V and 1.845 V vs RHE with the RBIP OFF and with the RBIP driven at duty cycles of 0.2 and 0.8. The frequency is 23 kHrz, the electrolyte was 2.75 m\textsc{m} \ce{HCl} aqueous solution, and all other parameters are as in Figure 5,.
 \supfigref[a]{CVshift} shows the measured curves. When driven at a duty cycle of 0.2, the HER onset  potential was shifted by 44 mV with respect to the HER onset when the ratchet is OFF.  Similarly, for \dc =0.8 HER onset was shifted by -65.9 mV with respect to the onset potential when the ratchet was OFF. 
Next, two more CV measurements were taken with the RBIP OFF. Here the potential range was modified according to the voltage shifts found above: the first was measured between potentials of -0.386 V and 1.801 V vs. RHE (a 44 mV cathodic change in range with respect to the original potential range),  and the second was measured between potentials of -0.276 V and 1.911 V vs. RHE (a 65.9 mV anodic change in range with respect to the original range). \supfigref[b]{CVshift} shows the measured curves and the curve measured initially with the ratchet OFF.
Last, each of the curves in \supfigref[b]{CVshift} was compared to the curve measured with the RBIP ON with the duty cycle that induced the corresponding potential shift. 
\supfigref[c]{CVshift}, shows  the CVs measured with RBIP driven with a duty cycle of 0.2 (solid line) and the CV measured with the RBIP OFF and the modified potential range (dashed line). To better compare the shapes of the two curves, the latter CV was  artificially shifted anodically by 44 mV by adding the potential shift to its potential data. Similarly, the CV measured with the RBIP driven at a duty cycle of 0.8 was compared to the CV measured with the RBIP OFF, and the CV potential range was augmented by -65.9 mV. Here, the  CV curve measured with the RBIP OFF was artificially shifted cathodically by 65.9 mV by subtracting 65.9 mV from its potential data. The comparison between these curves is shown in \supfigref[d]{CVshift}. The curves with the RBIP ON and the curves with the RBIP OFF and modified potential range show excellent agreement. Specifically, the OER and HER current onsets match almost perfectly (\supfigref[c,d]{CVshift}).  This demonstrates that in this electrode configuration, the RBIP functions as an additional voltage source, similar to an added potential induced by the potentiostat. If the RBIP functioned as a variable ionic resistance, it would have caused the slopes of the CV curve to change above the current onset and would not have shifted the curves as observed.

\supfig[\textwidth]{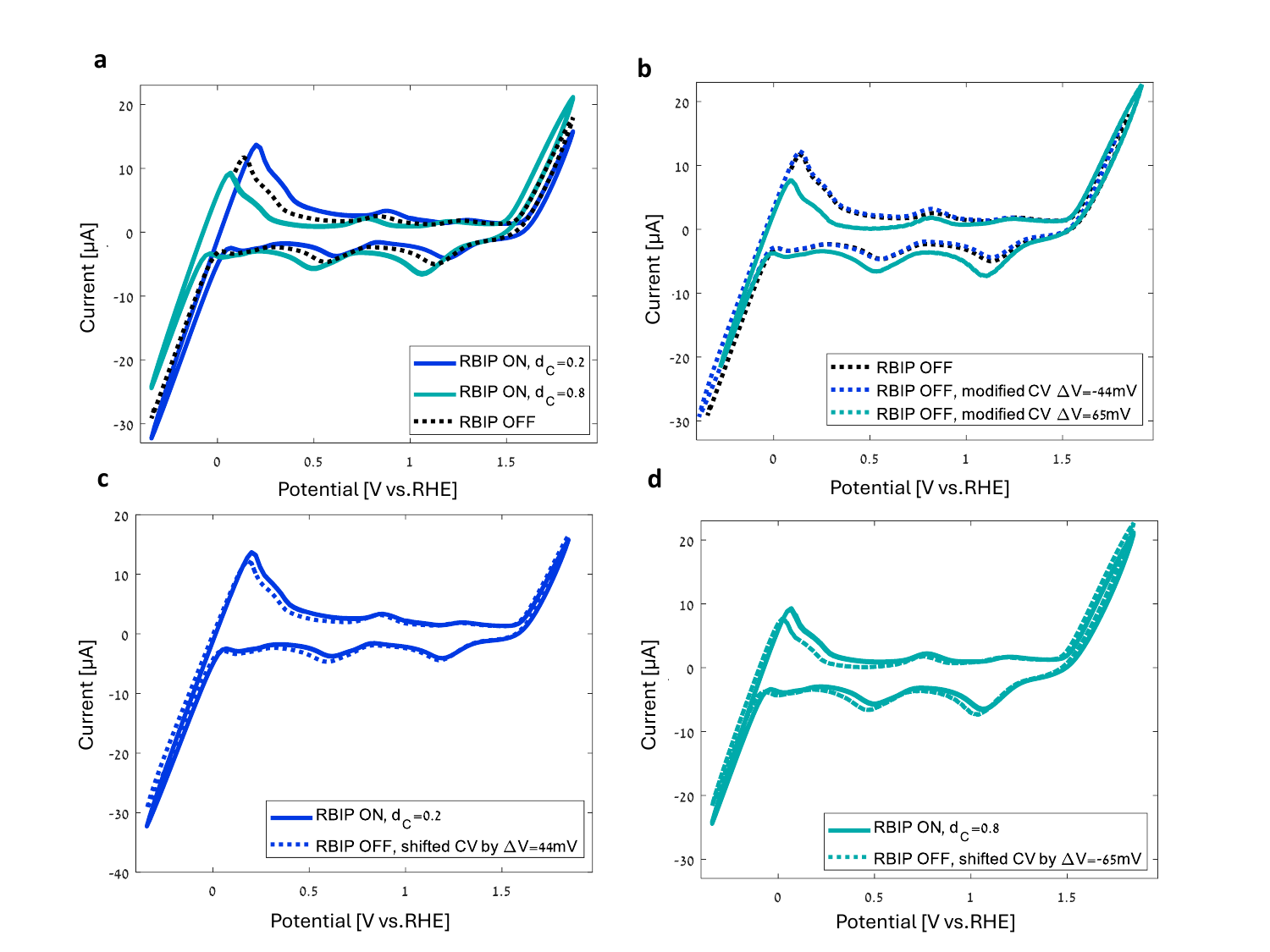}{(a) three electrode cyclic voltammetry measurement using a scan rate of $50 \space mVs^{-1}$ when the RBIP is off and when the RBIP is ON with  \dc =0.2 and \dc=0.8. The  frequency is 23 KHz and the amplitude is $V_{p-p}= 1.4\space V$. The RBIP was fabricated as described in the Experimental section with an alumina  ALD coating. (b)  CV curves measured with the RBIP OFF and the potential range modified by the potential shifts induced by the RBIP in (a). (c) A comparison between the  CV  curve measured in (a) for a duty cycle of 0.2  and the CV from (b) with the potential range modified by 44 mV. To better compare the shapes of the curves, the curve from (b) was artificially shifted cathodically by 44 mV by reducing 44 mV from its potential data. (d) A comparison between the  CV  curve measured in (a) for a duty cycle of 0.8  and the CV from (b) with the potential range modified by -65.9 mV. Here, the curve from (b) was artificially shifted anodically by 65.9 mV by adding 65.9 mV from its potential data.}{CVshift}
\pagebreak
\section{Cyclic voltammetry in configuration B}
 The measurements depicted in Figure 5 were repeated with the electrodes in configuration B. \supfigref{CVsweepcompare} shows a comparison between the CV curves measured in configurations A and B for various input signal duty cycles. \supfigref[a]{CVsweepcompare} shows the same data as in Figure 5a (configuration A), and \supfigref[d]{CVsweepcompare} shows the corresponding CV curves measured  in configuration B. \supfigref[b,e]{CVsweepcompare} shows respectively the RBIP induced currents  in configurations A and B, and \supfigref[c,f]{CVsweepcompare} shows respectively the HER onset and proton desorption potentials in configurations A and B.
When switching the electrode configuration, the shift in overpotential and current is reversed, revealing the ratchet directionality.
In configuration B, for duty cycles below 0.5, the  HER onset and proton desorption peak are shifted to more cathodic potentials (\supfigref[f]{CVsweepcompare}), leading to a current decrease (\supfigref[e]{CVsweepcompare}). At duty cycles above 0.5, the HER onset and proton desorption peak are shifted to more anodic potentials, resulting in an increase in current. Thus, as shown in Figure 3, the RBIP drives protons towards the $R^-$ compartment for duty cycles below 0.5 and towards the $R^+$ compartment for duty cycles above 0.5.
\supfig[\textwidth]{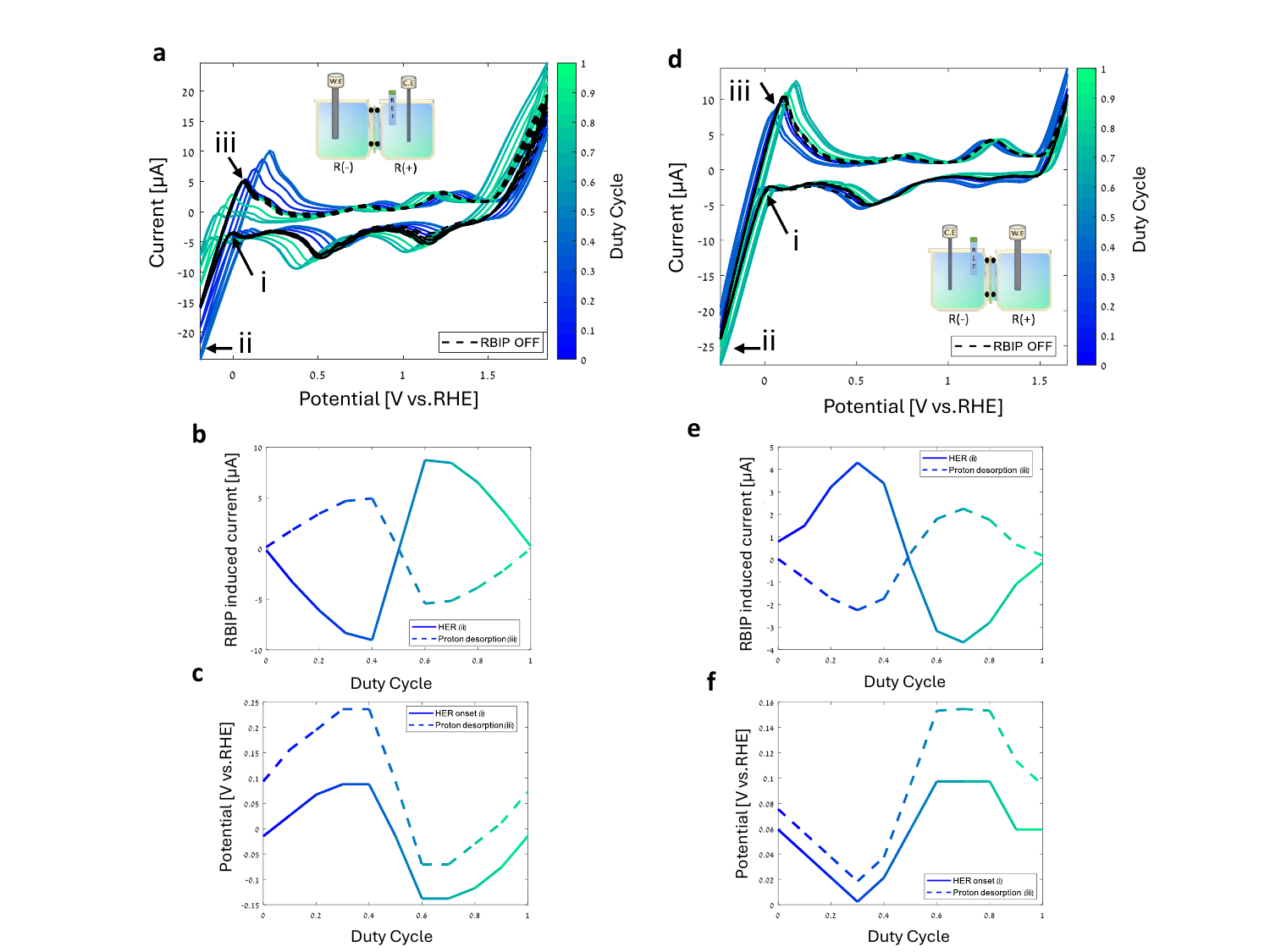}{
(a,d)  Three-electrode cyclic voltammetry curves measured in configurations A and B respectively, with the RBIP OFF ($V_{in}$= 0 V, dashed black curve) and with the RBIP driven at several input signal duty cycles. The scan rate is $50 \space mVs^{-1}$. The RBIP was fabricated as described in the Experimental section with an alumina ALD coating. The electrolyte is a 2.75 m\textsc{m} HCl aqueous solution.
The input signal frequency is 15 kHz, and the amplitude is $V_{p-p}= 1.4\space V$. (b) The  RBIP induced current as a function of the duty cycle  in configuration A extracted from (a). (c) The HER onset and proton desorption peak potentials as a function of the input signal duty cycle in configuration A as extracted from (a).  (e) The  RBIP induced current as a function of the duty cycle in configuration B as extracted from (d). (f) The HER onset and proton desorption peak potentials as a function of the input signal duty cycle in configuration B as extracted from (d).}
{CVsweepcompare}

\pagebreak
\section{Operation in sulfuric acid}  
The effect of the RBIP on the CV curves was also studied in sulfuric acid. The measurement procedure was as in Figure 5. \supfigref[a]{CVh2so4} shows the CV curves measured with the RBIP OFF ($V_{in}= 0 \space V$, dashed black curve) and with  input signals with various duty cycles applied to the RBIP. The scan rate is $50\space mVs^{-1}$ and the electrodes are in configuration A. The RBIP sample fabrication process was the same as used for the sample discussed in Figure 5, and the electrolyte is a 1.6 m\textsc{m}  \ce{H2SO4} aqueous solution. The input signal frequency is 23 kHz,  and the amplitude is $V_{p-p}= 1.4\space V$ . \supfigref[b]{CVh2so4} shows the HER onset and the proton desorption potentials as a function of the input signal duty cycle, and \supfigref[c]{CVh2so4} shows the RBIP-induced HER and proton desorption current change as a function of the input signal duty cycle. The CV curves show the same trends as the results in \ce{HCl} aqueous solution. The output is slightly higher than measured in \ce{HCl}  reaching a maximal HER current change of $-11.42\mu A$ and $11.62\mu A$ for duty cycles of 0.4 and 0.6, respectively, and an HER onset potential  shift of $175.9mV$ and $-153mV$ for duty cycles of 0.4 and 0.6, respectively.
\supfig[\textwidth]{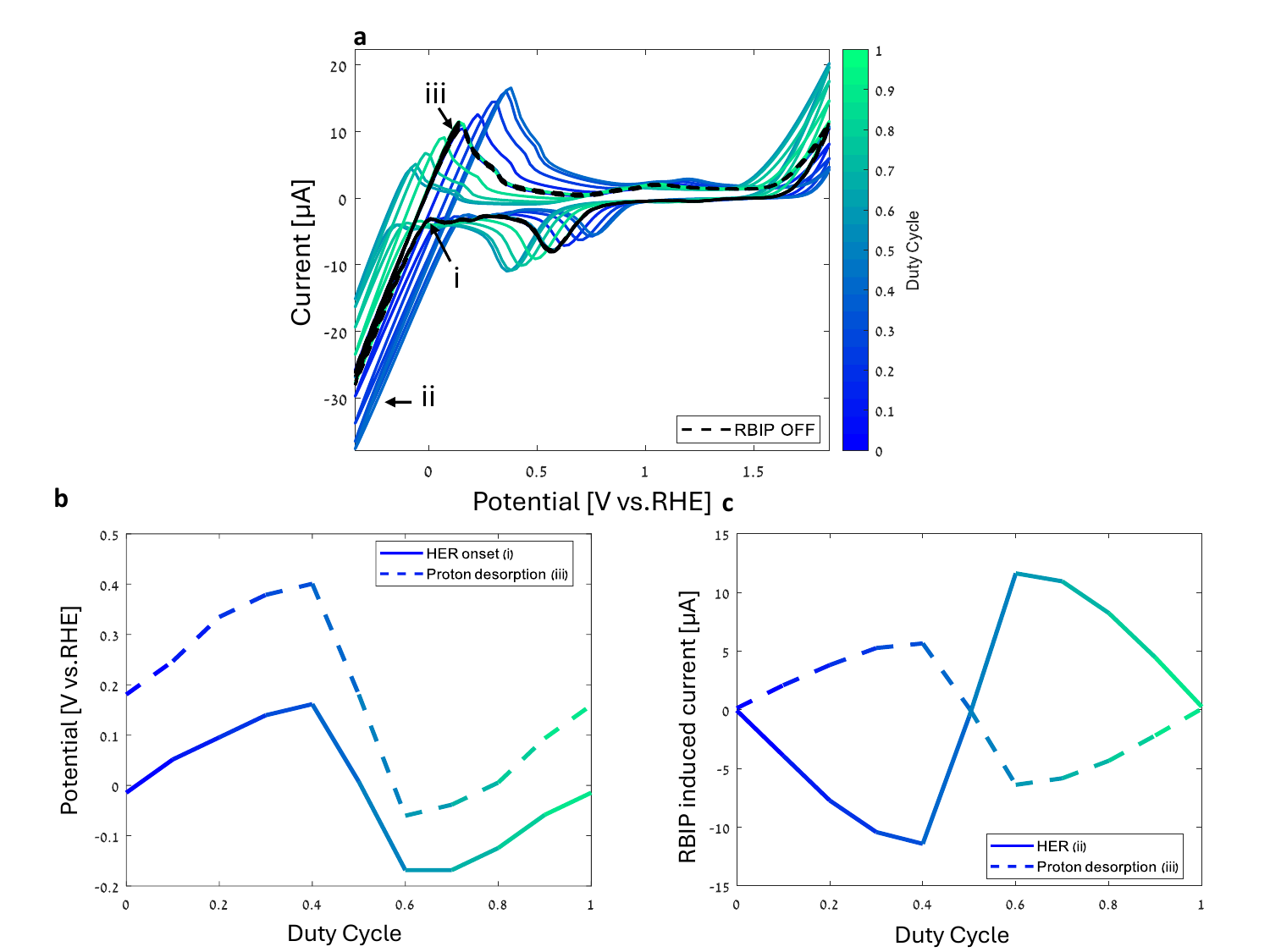}{(a) Three-electrode cyclic voltammetry measurements when the RBIP is OFF ($V_{in}$ = 0 V, dashed black curves) and with the RBIP driven with several input signal duty cycles. The scan rate is $50\space mVs^{-1}$. The RBIP was fabricated as described in the Experimental section with an alumina  ALD coating. The electrolyte was a 1.6 m\textsc{m}  \ce{H2SO4} aqueous solution. The input signal frequency was 23 KHz, and the amplitude was  $V_{p-p}= 1.4 \space V$ . (b) The extracted HER onset and proton desorption potentials as a function of the input signal duty cycle. (c) The extracted RBIP-induced HER and proton desorption current change as a function of the input signal duty cycle. }{CVh2so4}
\pagebreak
\section{Two-electrodes experiment }
To assure that the ion pumping effect is not an artifact introduced by electronic feedbacks in the potentiostat, the  CV measurements were also conducted in a 2-electrode configuration. The CV curve was measured once when the ratchet was OFF, and then was measured with the ratchet ON at duty cycles of 0.2 and 0.8. The input signal frequency was 10 kHz, and the amplitude was  $V_{p-p}= 1.4 \space V$. The electrolyte is a 1.6 m\textsc{m}  \ce{H2SO4} aqueous solution. Last, the CV curve was measured while a constant bias of -420 mV (the time-averaged voltage for an input signal with a duty cycle of 0.2)  was applied to the RBIP. The solution and sample were as in section 4 in the supplementary information.  \supfigref[]{CV2elec} shows the measured CV curves.  The trends are similar to the results observed in the 3-electrode experiments: when a square wave is applied, the CV shifts in potential according to the input signal duty cycle. However, a constant bias has no effect on the CV.  

\supfig[13cm]{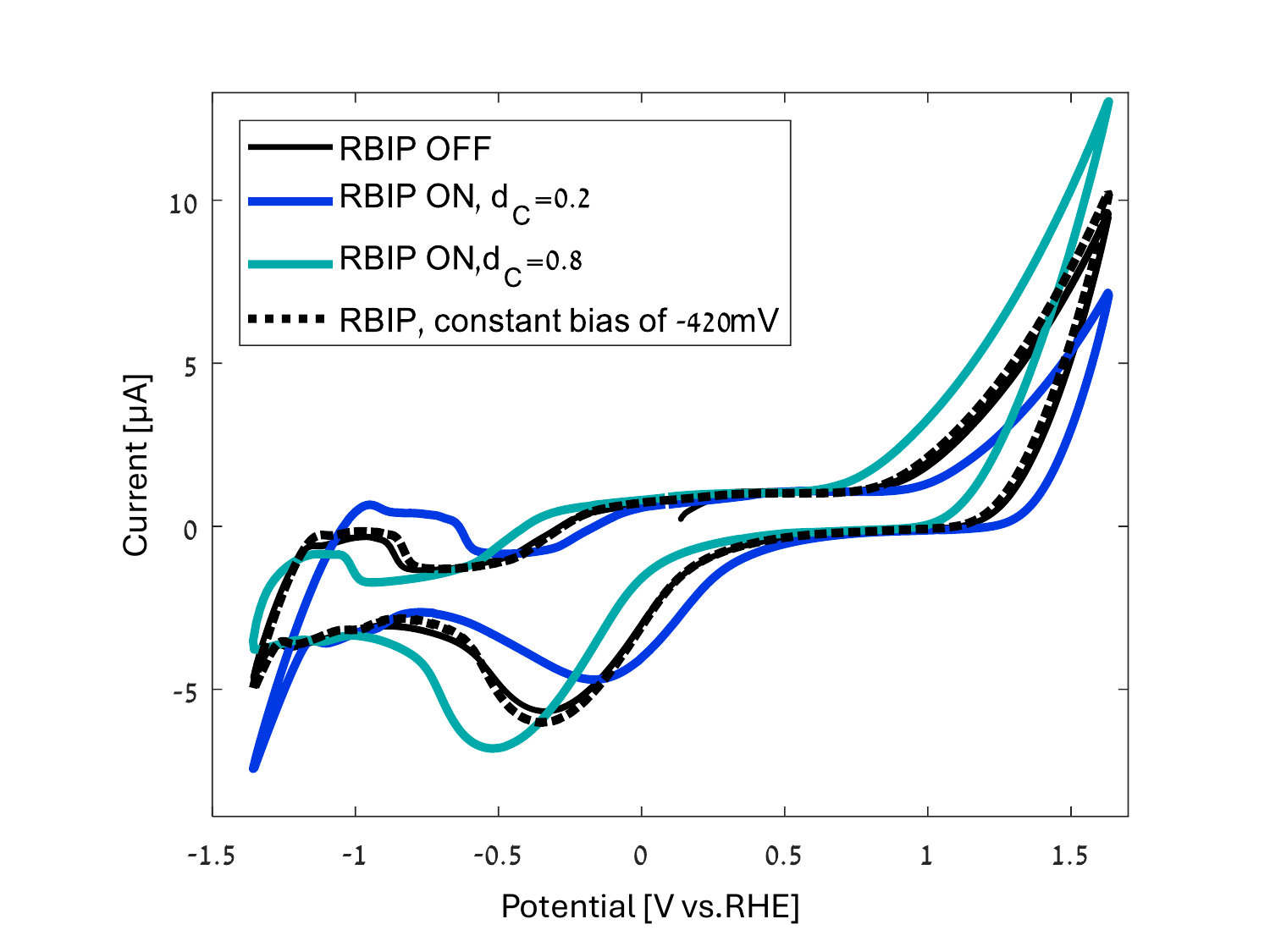}{Two-electrode cyclic voltammetry curves measured with the RBIP OFF ($V_{in }=0\space V$), driven with duty cycles of 0.2 and 0.8, and when a constant bias of -420 mV was applied to the RBIP. The input signal frequency was 10 kHz, and the amplitude was  $V_{p-p}= 1.4V$. The scan rate is $50\space mVs^{-1}$. The RBIP was fabricated as described in the Experimental section with an alumina  ALD coating. The electrolyte is a 1.6 m\textsc{m}  \ce{H2SO4} aqueous solution. }{CV2elec}
\supfig[13cm]{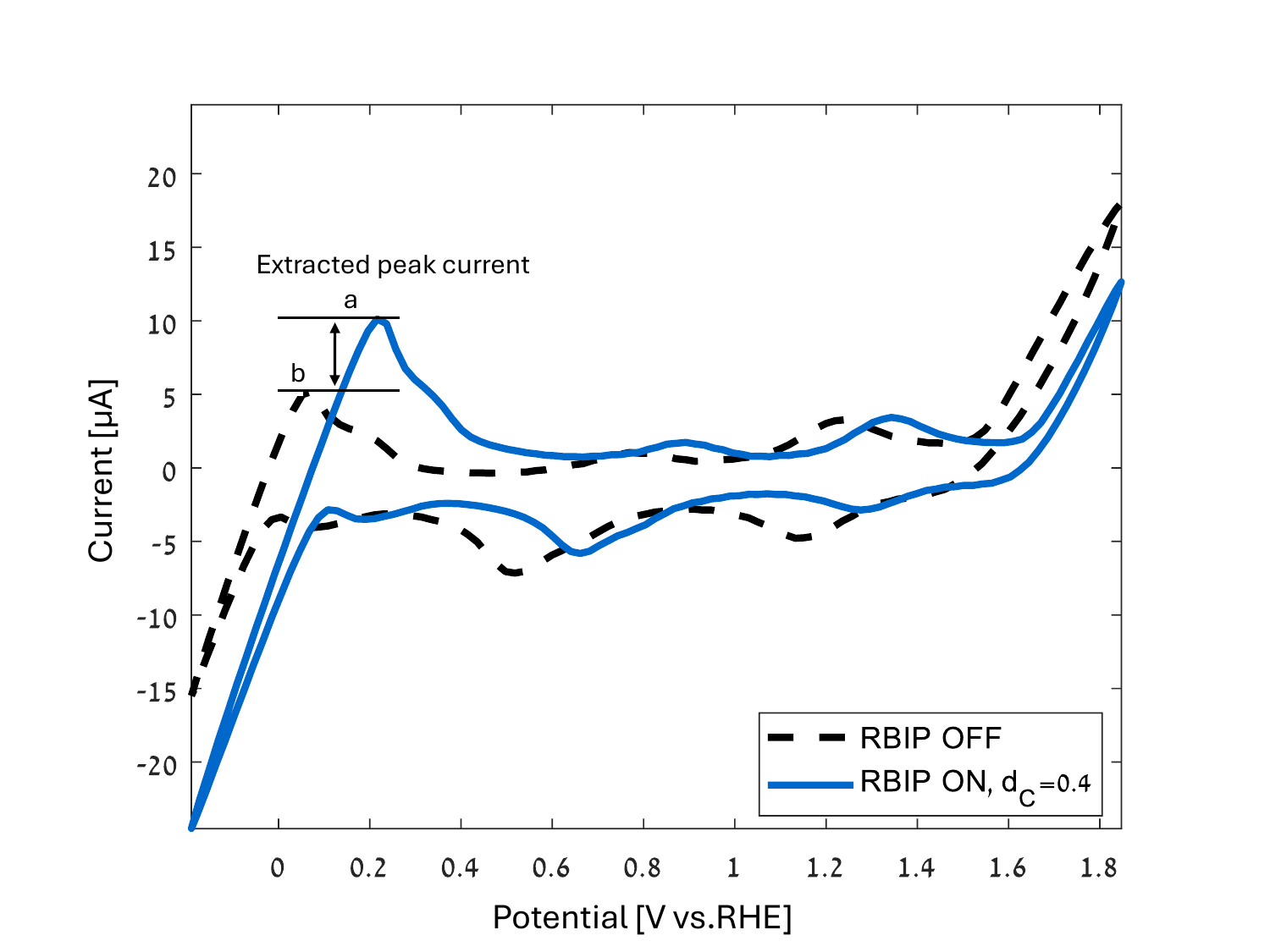}{graphical calculation example for the extracted RBIP induced current as discussed in Figure 5}{extcurr}